\documentclass[12pt,twoside,a4paper]{article}
\usepackage{amsmath,amssymb,latexsym,theorem,natbib,epsfig,color,subfigure}
\usepackage{multirow,graphicx,array,rotating}  
\usepackage{epstopdf,url,afterpage}
\usepackage{verbatim}
\usepackage{mathtools}
\usepackage[normalem]{ulem}
\usepackage{soul}
\newfont{\msa}{msam10 scaled\magstep1}
\newfont{\ssmsa}{msam9}

\def\crps{\mathop{\hbox{\rm CRPS}}}
\def\crpss{\mathop{\hbox{\rm CRPSS}}}

\def\bs{\mathop{\hbox {\rm BS}}}
\def\bss{\mathop{\hbox{\rm BSS}}}

\numberwithin{equation}{section}

\title{Parametric post-processing of dual-resolution precipitation forecasts}

\author{{\sc Marianna Szab\'o$^{1,2}$},  {\sc Est\'\i baliz Gasc\'on$^{3}$} and {\sc S\'andor Baran$^{1}$} \vspace*{0.5cm}\\
  $^1$Faculty of Informatics, University of Debrecen, Hungary\\
  $^2$Doctoral School of Informatics, University of Debrecen, Hungary\\
         $^3$European Centre for Medium-Range Weather Forecasts,\\
         Reading, United Kingdom
         }

         \date{}

\begin{document}
\pagestyle{myheadings}

\maketitle

\begin{abstract}
Recently, all major weather centres issue ensemble forecasts which even covering the same domain differ both in the ensemble size and spatial resolution. These two parameters highly determine both the forecast skill of the prediction and the computation cost. In the last few years, the plans of upgrading the configuration of the Integrated Forecast System of the European Centre for Medium-Range Weather Forecasts (ECMWF) from a single forecast with 9 km resolution and a 51-member ensemble with 18 km resolution
induced an extensive study of the forecast skill of both raw and post-processed dual-resolution predictions comprising ensemble members of different horizontal resolutions. 

We investigate the predictive performance of the censored shifted gamma (CSG) ensemble model output statistic (EMOS) approach for statistical post-processing with the help of dual-resolution 24h precipitation accumulation ensemble forecasts over Europe with various forecast horizons. As high-resolution, the operational 50-member ECMWF ensemble is considered, which is extended with a 200-member low-resolution (29-km grid) experimental forecast. The investigated dual-resolution combinations consist of (possibly empty) subsets of these two forecast ensembles with equal computational cost, being equivalent to the cost of the operational 50-member ECMWF ensemble. 

Our case study verifies that, compared with the raw ensemble combinations, EMOS post-processing results in a significant improvement in forecast skill and the differences between the various dual-resolution combinations are reduced to a non-significant level. Moreover, the semi-locally trained CSG EMOS is fully able to catch up with the state-of-the-art quantile mapping and provides an efficient alternative without requiring additional historical data essential in determining the quantile maps.

\end{abstract}

\section{Introduction}
  \label{sec1}
Capturing and modelling uncertainty is an essential need in any forecasting problem, and in weather prediction it may result in an enormous economical benefit. In the early 90's there was an important shift in the practice of weather forecasting from deterministic forecasts obtained using numerical weather prediction (NWP) models in the direction of probabilistic forecasting. The crucial step was the introduction of ensemble prediction systems (EPSs) in operational use in 1992 both at the  European Centre for Medium-Range Weather Forecasts (ECMWF) and the U.S. National Meteorological
Center. An EPS provides a range of forecasts corresponding to different runs of the NWP models, which are usually generated from random perturbations in the initial conditions and/or the stochastic physics parametrization. In the last decades, the ensemble method has become a widely used technique all over the world opening the door for probabilistic forecasting \citep{gr05}. 

Obviously, the larger the ensemble size, the higher the chance of correctly estimating forecast uncertainty \citep{ms16,l18}, while high spatial resolution is essential for detecting local phenomena. However, weather centres have a fixed amount of computational resources, and as computational costs increase both with ensemble size and with resolution, before introducing a new operational EPS configuration, a reasonable tradeoff on these key parameters should be made. At the moment the operational global 51-member medium-range (up to 15 days) ECMWF ensemble is generated at TCo639 resolution meaning an approximately 18 km grid spacing \citep{ecmwfEval18}, whereas for 2023 the plan is to reach the 9 km (TCo1279) resolution of the current deterministic forecast. The new configuration to be implemented in 2023 will comprise a 51-member medium-range TCo1279 forecast run twice a day and a 101-member extended-range (up to 46 days) forecast on a 36 km grid (TCo319) run only once daily. Though the change in configuration was driven by the separate benefits in skill for the medium-range ensemble originating from the resolution increase and the benefit in skill for the extended-range forecasts from the increase in ensemble size, this new setup will allow operational dual-resolution ensemble forecasts. Further, according to the ECMWF Strategy for 2021--2030\footnote{\url{https://www.ecmwf.int/sites/default/files/elibrary/2021/ecmwf-strategy-2021-2030-en.pdf} [Accessed on 23 December 2022]}, "ECMWF will continue to investigate a mixture of larger ensemble and increased vertical and horizontal resolution, and a blend of variational and ensemble methods across the Earth system components". 
In line with the above goals, \citet{lbb20} investigated the forecast skill of dual-resolution ensembles by considering combinations of high- (TCo639) and low resolution (TCo399, grid resolution $\approx$ 29 km; TCo255, grid resolution $\approx$ 45 km) temperature predictions for a given computation costs. The authors found that in the case of 2m temperature, provided the ensemble size is large enough, combinations with roughly equal number of high- and low resolution members exhibit the best predictive performance.

However, ensemble forecasts are often underdispersive, that is, the spread of the ensemble is too small to account for the full uncertainty, and may also be subject to systematic bias. This phenomenon has been observed with several operational EPSs \citep[see e.g.][]{bhtp05} and can be resolved by some form of statistical post-processing \citep{buizza18}. Over the last decades, various post-processing methods have been proposed for a large variety of weather variables; for a detailed overview of the most advanced methods see e.g. \citet{w18} or \citet{vbde21}. Parametric approaches such as Bayesian model averaging \citep{rgbp05} or ensemble model output statistics \citep[EMOS:][]{grwg05} provide full predictive distributions of the future weather quantities, non-parametric methods usually capture predictive distributions via estimating their quantiles \citep[see e.g.][]{fh07,brem19}, whereas member-by-member post-processing uses e.g. linear regression to improve the raw ensemble \citep{vsv15}. Recently approaches using artificial neural networks also become more and more popular as they provide more flexibility in modelling both in parametric \citep[see e.g.][]{rl18,sswh20,gzshf22} and non-parametric context \citep{brem20}.

Concerning dual-resolution ensemble forecasts, \citet{blszbb19} investigated whether statistical post-processing changed the conclusions regarding the optimal dual-resolution configuration found by \citet{lbb20}. The authors found that EMOS calibration strongly reduced the differences in skill among the equal-cost configurations of single- and dual-resolution 2m temperature ensembles and the ranking of the different configurations could also change. In a parallel study, \citet{esti19} studied post-processing of ECMWF dual-resolution (TCo639 and TCo399) precipitation ensemble forecasts using a quantile mapping and objective weighting of sorted ensemble members approach \citep{hs18} and confirmed the superiority of the combination with equal number of high- and low resolution ensemble members. However, the considered calibration method requires an extended set of historical data; in \citet{esti19} forecasts-analysis pairs for 20 years were considered. In this paper, we consider the essentially simpler censored and shifted gamma (CSG) distribution-based EMOS approach of \citet{bn16} to calibrate ECMWF dual-resolution forecasts and compare the results with the findings of \citet{esti19}. The advantage of this parametric model is that it requires no additional historical data and results in full predictive distributions.

The paper is organized as follows. Section \ref{sec2} contains a detailed description of the studied precipitation accumulation dataset. In Section \ref{sec3} the applied calibration approaches are reviewed and the parameter estimation methods and model verification tools are provided. The results are provided in Section \ref{sec4}, followed by a concluding Section \ref{sec5}.

\section{Data}
 \label{sec2}
 
As it was established in the Introduction, the datasets used in this study are identical to the ones considered in \citet{esti19}. The weather variable at hand is 24h precipitation accumulation (from 0600 UTC to the same time of the next day) and the dual-resolution system consists of different combinations between ensemble forecasts at resolutions TCo639 (high resolution) and TCo399 (low resolution). Note that the cost ratio between these two resolutions is 4:1, so in the different dual-resolution configurations four TCo399 members can be traded against a single TCo639 forecast. 

\begin{figure}[h!]
\centering
\hbox{
\includegraphics[width=\textwidth]{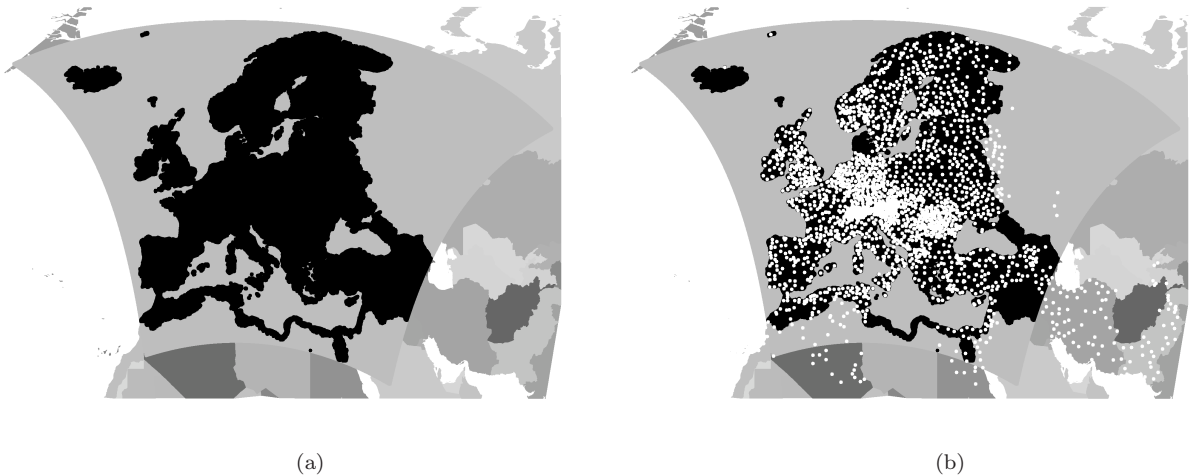}
}
    \centerline{\hbox to 10cm{\scriptsize \qquad\qquad (a) \hfill  (b)}} 
     \caption{Map of the (a) domain of the EFAS gridded data and the land subset; (b) SYNOP stations in the land subset of the EFAS gridded data.}
   \label{fig:mapsynop}
\end{figure}

The first dataset consists of 24h gridded accumulated precipitation analyses of the European Flood Awareness System \citep[EFAS:][]{nefas} for 1996--2016 covering Europe and some of the surrounding countries (see Figure \ref{fig:mapsynop}a). Data of 2016 serve as validation data for the investigated post-processing approaches, whereas analyses from the preceding years (1996--2015) are required for training the quantile mapping-based methods. Note that whether model training is performed using all EFAS grid points corresponding to the land subset (5 km grid spacing, 363 534 grid points), for validation purposes only data of 2370 grid points corresponding to SYNOP stations are considered (Figure \ref{fig:mapsynop}b). 

Post-processing is applied to 24h precipitation accumulation forecasts of the ECMWF's Integrated Forecast System (IFS) for June-July-August (JJA) 2016 with forecast horizons up to 10 days. All forecasts are initialized at 0000 UTC and in order to match the accumulation period of the EFAS analyses, lead times of 6h, 30h, \ldots , 246h are considered. Similar to \citet{esti19,blszbb19} and \citet{lbb20}, 50 perturbed members of the operational TCo639 ensemble and forecasts from the 200-member TCo399 experiment are analyzed. The investigated dual-resolution mixtures \ $(M_H,M_L)$ \ of \ $M_H$ \ high-resolution and \ $M_L$ \ low-resolution members are $$(50,0), \ (40,40), \ (20,120), \ (10,160), \ (0,200),$$ all having the same computational cost corresponding to the available HPC resources of the ECMWF at the moment the forecasts were generated.

Finally, quantile mapping-based approaches are trained with the help of 11-member gridded reforecasts for JJA of the period 1996--2016 with forecast horizons matching the lead times of the dual-resolution combinations generated both at TCo639 and TCo399 resolutions. For a more detailed description on the above datasets we refer to \citet{esti19} and the references therein.

\section{Statistical post-processing}
    \label{sec3}

In terms of computational costs and model complexity, EMOS is one of the most efficient post-processing approaches \citep[see e.g.][Fig. 1]{vbde21} showing excellent performance for a large variety of weather quantities. It fits a single parametric law to the ensemble forecast with parameters depending on the ensemble members. EMOS models for different weather variables usually differ in the applied parametric law and/or in the link functions connecting the ensemble forecasts to the parameters of the predictive distribution.

\subsection{Censored and shifted gamma EMOS model}
  \label{subs3.1}

In contrast to continuous weather quantities, such as temperature or wind speed that can be modelled with a normal \citep{grwg05} or a truncated normal distribution \citep{tg10}, respectively, precipitation accumulation has a special discrete-continuous nature. Hence, one can only consider non-negative predictive distributions assigning positive mass to the zero precipitation event. A popular choice is to consider a continuous distribution that can take both positive and negative values and left censor it at zero \citep{sch14,sh15}, which approach is also the core idea of the CSG distribution-based EMOS model suggested by \citet{bn16}.

Let \ $G_{\kappa,\theta}$ \ denote the cumulative distribution function (CDF) of a gamma distribution \ $\Gamma(\kappa,\theta)$ \ with shape \ $\kappa>0$ \ and \ scale \ $\theta>0$ \ defined by probability density function (PDF)
\begin{equation*}
g_{\kappa,\theta}(x)\coloneqq 
\begin{cases}
       \frac{x^{\kappa -1}{\mathrm e}^{-x/\theta}}{\theta^{\kappa}\Gamma(\kappa)}, &\qquad x>0,\\
       0, &\qquad\text{otherwise,} \\
     \end{cases}
\end{equation*}
where \ $\Gamma(\kappa)$ \ stands for the value of the gamma function at \ $\kappa$. \ Since there is a one-to-one correspondence between the parameters \ $\kappa$ \ and \ $\theta$ \ and the mean \ $\mu >0$ \ and standard deviation \ $\sigma >0$ \ of the corresponding gamma distribution via equations  
\begin{equation*}
\kappa = \frac{\mu^2}{\sigma^2} \qquad \text{and} \qquad \theta = \frac{\sigma^2}{\mu}, 
\end{equation*}
$\Gamma(\kappa,\theta)$ \ can be characterized by these two moments as well. After extending the support of the gamma distribution to negative values with the help of a shift parameter \ $\delta >0$, \ one can introduce a shifted gamma distribution, left censored at zero \ $\Gamma^0(\kappa,\theta,\delta)$ \ with shape \ $\kappa$, \ scale \ $\theta$ \ and shift \ $\delta$ \ by CDF 
\begin{equation*}
G^0_{\kappa,\theta,\delta}(x)\coloneqq 
\begin{cases}
       G_{\kappa,\theta}(x+\delta), &\quad x\geq0,\\
       0, &\quad x<0. \\
     \end{cases}
\end{equation*}

In what follows, let \ $f_1, f_2 ,\ldots , f_K$ \ denote the precipitation accumulation ensemble forecast for a given location, time and lead time, and denote by \ $\overline f$ \ the ensemble mean. In the CSG EMOS model mean \ $\mu$ \ and variance \ $\sigma^2$ \ of the underlying gamma distribution are linked to the ensemble members as
\begin{equation}
    \label{CSGemos}
    \mu = a^2 + b_1^2 f_1 + b_2^2 f_2 + \cdots + b_K^2 f_K \qquad \text{and} \qquad \sigma^2 = c^2 + d^2 \overline f,
\end{equation}
whereas shift parameter \ $\delta >0$ \ is independent of the ensemble forecast.

However, the 50 perturbed members of the operational ECMWF IFS are considered as statistically indistinguishable and, in this way, exchangeable, and the same applies to the investigated 200-member low-resolution ensemble as well. In this way the investigated dual-resolution forecasts consists of groups of exchangeable ensemble members, which should be taken into account in the modelling process. In the following, if the $M$-member ensemble is divided into \ $K$ \ groups of exchangeable forecasts, where the $k$th group contains \ $M_k \geq 1$ \ ensemble members \ ($\sum_{k=1}^M M_k = M$), \ notation \ $\overline f_k$ \ will be used for the corresponding group mean. In this situation, ensemble members within a given exchangeable group should share the same parameters \citep{w18} and link functions in \eqref{CSGemos} should be replaced by
\begin{equation}
    \label{CSGemosEx}
    \mu = a^2 + b_1^2 \overline f_1 + b_2^2 \overline f_2 + \cdots + b_K^2 \overline f_K \qquad \text{and} \qquad \sigma^2 = c^2 + d^2 \overline f.
\end{equation}

\subsection{Parameter estimation}
\label{subs3.2}
Following the optimum score estimation principle of \citet{gr07}, mean parameters \ $a,b_1,\ldots, b_K$, \ variance parameters \ $c,d$, \ and shift parameter \ $\delta >0$ \  of the CSG EMOS model specified either by \eqref{CSGemos} or by \eqref{CSGemosEx} can be estimated by optimizing the mean value of a proper scoring rule over an appropriate set of training data. Scoring rules measure predictive performance by assigning numerical values to forecast-observation pairs and in atmospheric sciences one of the most popular choices is the continuous ranked probability score \citep[CRPS:][Section 9.5.1]{w19}, as it assesses simultaneously both calibration and sharpness of the probabilistic forecast. For a prediction represented by a CDF \ $F(y)$ \ and a real value \ $x$ \ the CRPS is defined as
\begin{equation}
\label{crps}
\crps(F,x) \coloneqq \int_{-\infty}^{\infty}\left[F(y)-\mathbb{I}_{\{y\geq x\}}\right]^2{\mathrm d}y,
\end{equation}
with \ $\mathbb{I}_H$ \ denoting the indicator function of a set \ $H$. \ Note that CRPS can be reported in the same units as the observation and it is a negative oriented score where smaller values mean better forecast skill. Further, for the CSG distribution CRPS can be expressed in a closed form \citep{sh15} allowing for efficient optimization in the parameter estimation process.

The next step in EMOS modelling is the appropriate choice of training data. The most common approach is the use of rolling training periods where ensemble forecasts and corresponding validation observations from the preceding \ $n$ \ calendar days are considered. The spatial selection of the training data is also an important issue, where the traditional approaches are local and global (regional) modelling \citep{tg10}. In the former case EMOS model parameters for a given location are estimated using training data of that particular spot, whereas in the latter all available forecast cases in the training data are considered resulting in a single set of parameters for the whole ensemble domain. In general, local EMOS models outperform their regional counterparts; however, in order to avoid numerical problems in parameter estimation, they require rather long training periods \citep{hspbh14}. 
In contrast, regional modelling can be performed using much shorter training periods \citep[see e.g.][]{bb21}, though it is usually unsuitable for large and heterogeneous ensemble domains like the one at hand (Figure \ref{fig:mapsynop}a). The advantages of the above selection methods can be combined using semi-local approaches, where one either augments the training data for a given location with data of locations with similar characteristics, or splits the ensemble domain into more homogeneous subdomains and within each subdomain performs a regional modelling \citep{hhw08,lb17}. The data augmentation technique of \citet{hs18} is applied in the reference quantile mapping approaches described in Section \ref{subs3.3}, whereas in EMOS modelling we consider the clustering-based semi-local method of \citet{lb17}. For each date in the verification period subdomains are formed dynamically by grouping the EFAS grid points of the land subset into clusters using $k$-means clustering of feature vectors depending both on the grid point climatology and the forecast errors of the raw ensemble during the training period.

\subsection{Quantile mapping}
 \label{subs3.3}

In the case study of Section \ref{sec4}, the reference post-processing approach is quantile mapping and its weighted version investigated by \citet{esti19}. The main idea behind the quantile mapping is that using climatological CDFs \ $F_f(y)$ \ and \ $F_o(y)$ \ of forecasts and observations, respectively, a raw forecast \ $f$ \ is adjusted to match the distribution of the observation \ $x$. \ The adjusted forecast \ $\widetilde f$ \ is given by
\begin{equation}
    \label{qmformula}
    \widetilde f \coloneqq F_o^{-1}\big(F_f(f)\big),
\end{equation}
and for a forecast ensemble the adjustment should be made separately for each member. 

In our case \ $F_f(y)$ \ and \ $F_o(y)$ \ are estimated from historical data of calendar years 1996 -- 2015 using the control of the 11-member ECMWF reforecasts and the EFAS analysis, respectively. For each grid point and each date of the verification period (JJA 2016) climatological CDFs are developed from 9000 sample values corresponding to the 9 closest Julian dates to the given date from the whole 20-year period for 50 supplemental similar locations chosen according to suggestions of \citet{hs18}.

For the weighted version of quantile mapping first each member of the ECMWF reforecast for 1996 -- 2015 has to be adjusted separately. For a given calendar year, climatological CDFs are calculated in the same way as before utilizing the matching reforecasts and corresponding analyses of 9 neighbouring dates from the remaining 19 years for 50 similar supplemental locations \citep[for the details see ][Section 2.4.1]{esti19}.

The 11-member quantile-mapped reforecasts for the period from 1996 to 2015 are then applied to derive the 11-bin closest-member histograms, that is histograms of ranks of the adjusted reforecast members nearest to the analysed precipitation amount, for various quantile-mapped ensemble mean values. By fitting a beta distribution to an 11-bin closest member histogram one can generate weights either for the operational TCo639 or for the experimental TCo399 ensemble forecasts (controls included, resulting in 51 and 201 bins, respectively). For a detailed description of weighted quantile-mapping we again refer to \citet{hs18} and \citet{esti19}. Dual-resolution combinations are then formed from the weighted members of TCo639 and TCo399 forecasts.
 
\subsection{Verification scores}

The predictive performance of various probabilistic forecasts is quantified with the help of the mean CRPS over all forecast cases in the verification period. Besides this verification measure we also consider the Brier score \citep[BS:][Section 9.4.2]{w19} for the dichotomous event that the observed precipitation accumulation \ $x$ \ is above a particular threshold \ $y$. \ Given again a predictive CDF \ $F(y)$ \ representing a probabilistic  forecasts, the BS  is defined as
\begin{equation}
    \label{bs}
\bs(F,x;y) \coloneqq \big( F(y) - \mathbb{I}_{\{y\geq x\}} \big)^2.
\end{equation}
In line with the reference study of \citet{esti19}, in Section \ref{sec4} we report the mean BS of 24h accumulated precipitation for thresholds  $0.1$,  $5$ and $10$ mm. Note that $\bs$ is again a negatively oriented score and the $\crps$ is the integral of the $\bs$ over all thresholds. Further, for a probabilistic forecast provided in the form of a forecast ensemble, both in \eqref{crps} and in \eqref{bs} the predictive CDF \ $F$ \ should be replaced by the empirical one.

One can gain a better insight into the smaller differences in the predictive performance of the competing forecasts by examining the continuous ranked probability skill scores \citep[CRPSS: see e.g.][]{gr07} and Brier skill scores \citep[BSS: see e.g.][]{ft12} quantifying improvement in CRPS and BS of a forecast \ $F$ \ over a reference forecast \ $F_{ref}$, \ respectively. If \ $\overline{\crps}, \ \overline{\bs}$ \ and \ $\overline{\crps}_{ref}, \ \overline{\bs}_{ref}$ \ denote the mean score values over the verification data corresponding to \ $F$ \ and \ $F_{ref}$, \ respectively, \ $\crpss$  and $\bss$ are defined as
\begin{equation*}
\crpss \coloneqq 1 - \frac{\overline{\crps}}{\overline{\crps}_{ref}} \qquad \text{and} \qquad \bss \coloneqq 1 - \frac{\overline{\bs}}{\overline{\bs}_{ref}}.
\end{equation*}

The calibration of probabilities of a dichotomous event of exceeding a given threshold calculated from the various competing forecasts is compared with the help of reliability diagrams \citep[][Section 9.4.4]{w19} depicting the graph of the observed relative frequencies of the event against the corresponding binned forecast probabilities. In the case of proper calibration this graph should lie on the main diagonal of the unit square. We follow the suggestions of \citet{bs07} and plot the observed relative frequency of a bin against the mean of the corresponding probabilities and we also add inset plots showing the frequencies of the bins on $\log10$ scales.

Finally, the statistical significance of the differences between the verification scores is assessed with two different methods. From the one hand, we report confidence intervals for the mean score values and skill scores calculated from 2 000 block bootstrap samples based on the stationary bootstrap scheme with mean block length according to \citet{pr94}. From the other hand, we apply the Diebold–Mariano (DM) test for equal predictive performance \citep{dm95}, which is able to account for temporal dependencies. In simultaneous testing for the different locations, we also handle spatial dependencies by applying a Benjamini-Hochberg algorithm \citep{bh95} to control the false discovery rate at a 5\,\% level of significance \citep[see e.g.][]{w16}.

\section{Results}
    \label{sec4}
    
As mentioned in the Introduction, in this study we investigate the effect of CSG EMOS post-processing (see Section \ref{subs3.1}) on various combinations of TCo639 and TCo399 24h precipitation accumulation ensemble forecasts, including the pure high- and pure low-resolution ensemble. In the following analysis, notation \ $f_{H,1},f_{H,2}, \ldots , f_{H,M_H}$ \ is used for the ensemble members at TCo639 resolution and \ $f_{L,1},f_{L,2}, \ldots , f_{L,M_L}$ \  for the TCo399 members of the dual-resolution forecast for a given location, time point and lead time. As for both resolutions we consider only forecasts obtained using perturbed initial conditions, ensemble members at a given resolution can be considered as exchangeable. Hence, link functions \eqref{CSGemosEx} of the CSG EMOS model reduce to
\begin{equation}
    \label{CSGemosDR}
    \mu = a^2 + b_H^2 \overline f_H + b_L^2 \overline f_L \qquad \text{and} \qquad \sigma^2 = c^2 + d^2 \overline f, 
\end{equation}
where \ $\overline f_H$ \ and  \ $\overline f_L$ \ denote the mean of high- and low-resolution members, respectively. Model parameters are estimated by minimizing the mean CRPS over the training data where we fix  \ $b_L=0$ \ in the pure high-resolution  \ ($M_L=0)$ \ and \ $b_H=0$ \ in the pure low-resolution  \ ($M_H=0)$ \  case.

Due to the large number of zeros both in predicted and observed precipitation accumulation, statistical post-processing of this weather variable requires far more training data than e.g. temperature or wind speed. For local EMOS models \citet{hspbh14} suggests to use data of almost 5 years (1816 calendar days), whereas our ECMWF dual-resolution forecasts cover just the 97-day time interval between 1 June 2016 to 5 September 2016. The extension and heterogeneity of the EFAS domain (Figure \ref{fig:mapsynop}a) makes regional modelling unreliable as well, so as mentioned in Section \ref{subs3.2}, a clustering-based semi-local approach is applied. After a detailed data analysis where several combinations of training period length and number of clusters had been tested, we decided to estimate parameters of the CSG EMOS model over 8000 clusters using a 30-day rolling training period. Similar to \citet{lb17}, clustering is performed using 24-dimensional feature vectors, where half of the features are obtained by taking equidistant quantiles of the climatological CDF over the training period, whereas the other half consists of the same quantiles of the empirical distribution of the forecast error of the ensemble mean. This configuration results in on average 1363 forecast-observation pairs for each estimation task (5 or 6 parameters to be estimated) and leaves 52 calendar days verification purposes (period 11 July 2016 -- 31 August 2016). As noted before, CSG EMOS post-processed predictions are validated using data of just 2370 SYNOP stations (Figure \ref{fig:mapsynop}b), allowing a direct comparison with quantile-mapped (QM) and weighted quantile-mapped (QM+W) forecasts of \citet{esti19}, and to be fully in line with this work, we report the various verification scores only for forecast horizons 1, 3, 5, 7 and 10 days.

\begin{figure}[t!]
\centering
\includegraphics[width=0.7\textwidth]{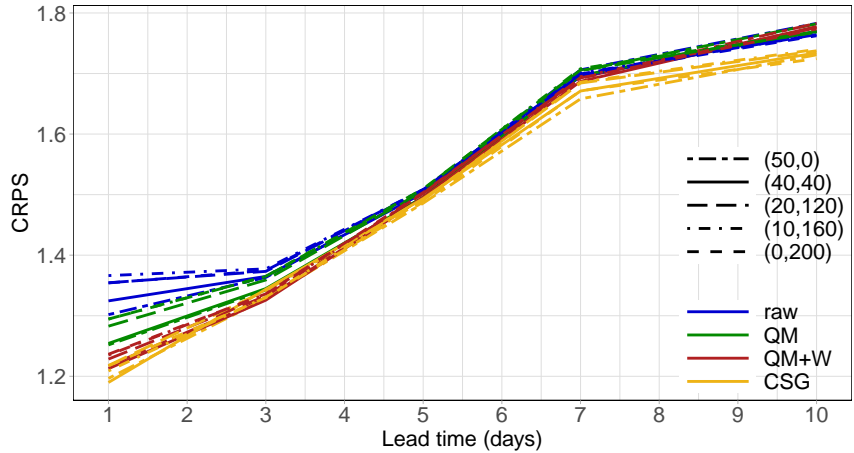}
     \caption{CRPS of raw and post-processed forecasts as function of the forecast horizon.}
   \label{fig:crps}
\end{figure}

Figure \ref{fig:crps} shows the mean CRPS of raw and post-processed forecasts of the investigated dual-resolution combinations. Up to day 5 all post-processed forecast combinations outperform all raw dual-resolution forecasts, the largest differences appear at day 1, whereas the smallest gain appears at day 5. For longer lead times the advantage of QM and QM+W forecasts disappears and the CSG EMOS models result in the lowest mean CRPS. 

\begin{figure}[t!]
\centering
\hbox{
\includegraphics[width=\textwidth]{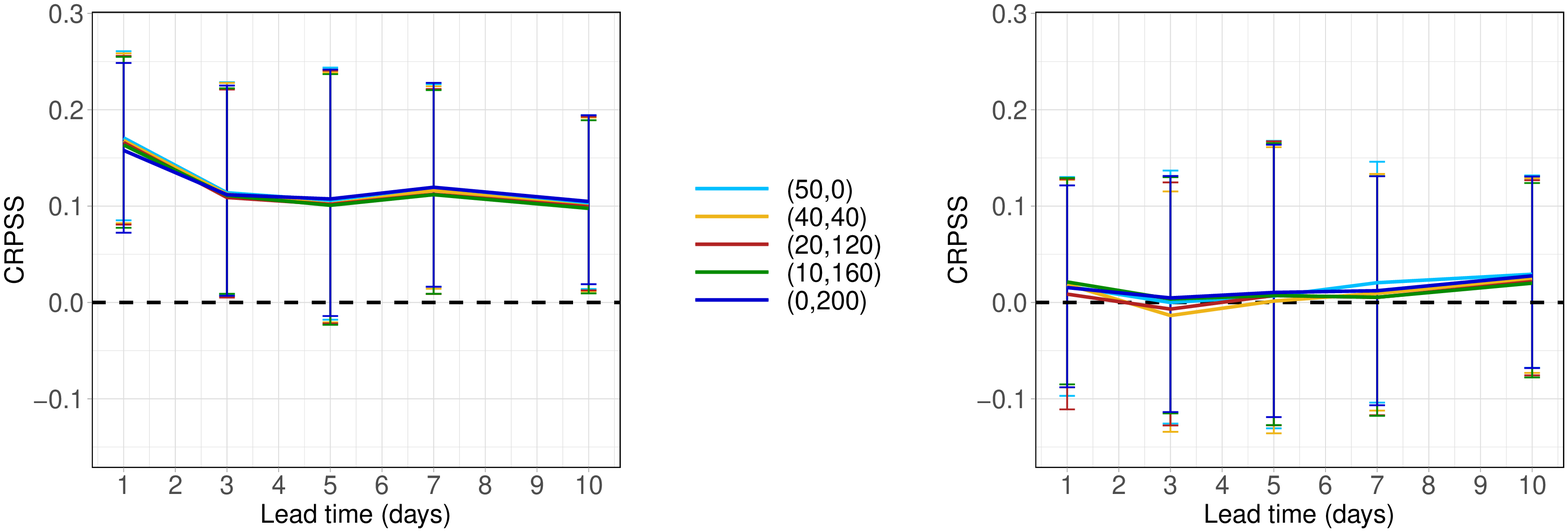}
}
    \centerline{\hbox to 12cm{\scriptsize \qquad (a) \hfill (b)}} 
     \caption{CRPSS of the CSG EMOS model for different dual-resolution configurations (a) with respect to the raw (50,0) combination; (b) with respect to the corresponding QM+W forecast with 95\,\% confidence intervals.}
   \label{fig:crpss}
\end{figure}

A slightly better insight into the differences between the various forecasts can be obtained from Figure \ref{fig:crpss}, where the skill scores of the dual-resolution CSG EMOS models with respect to the raw pure high-resolution (50,0) forecast (Figure \ref{fig:crpss}a) and with respect to the corresponding QM+W forecast (Figure \ref{fig:crpss}b) are plotted.  The difference in skill between CSG EMOS models corresponding to various dual-resolution combinations are minor, which is in line with the findings of \citet{blszbb19}, and they significantly outperform the raw high-resolution forecasts for all investigated lead times except day 5. Further, according to Figure \ref{fig:crpss}b, the essentially simpler CSG EMOS approach is completely able to catch up with the QM+W forecasts for all dual-resolution configurations and all lead times.
\begin{figure}[t!]
\centering
\hbox{
\includegraphics[width=\textwidth]{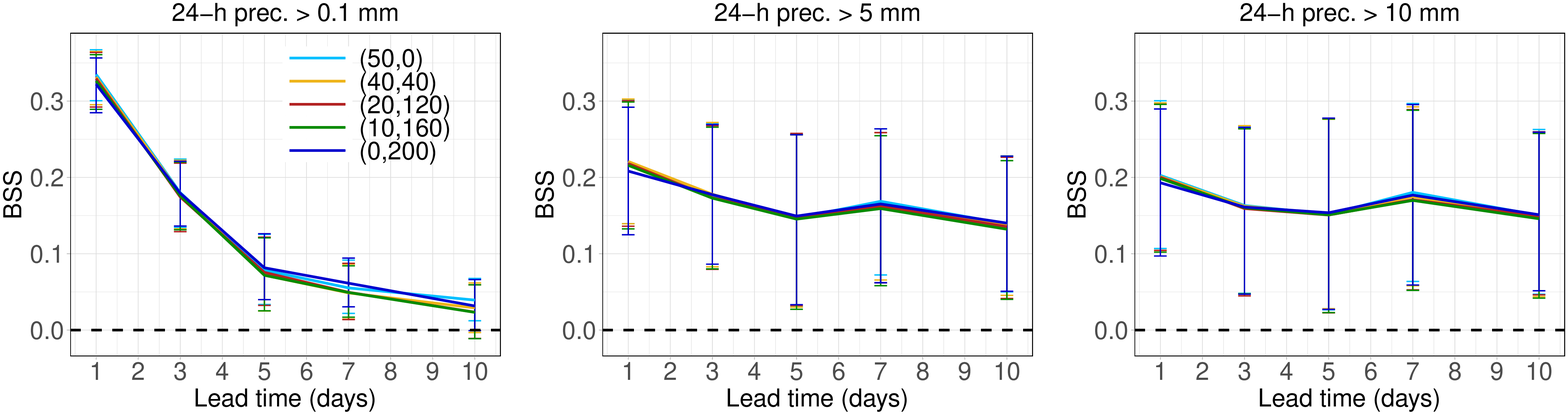}
}
    \centerline{\hbox to 12cm{\scriptsize \qquad (a) \hfill  (b) \hfill (c)}} 
    \caption{BSS of the CSG EMOS model for different dual-resolution configurations with respect to the raw (50,0) configuration with 95\,\% confidence intervals for thresholds (a) 0.1 mm; (b) 5 mm; (c) 10 mm.}
   \label{fig:bssraw}
\end{figure}

\begin{figure}[t!]
\centering
\hbox{
\includegraphics[width=\textwidth]{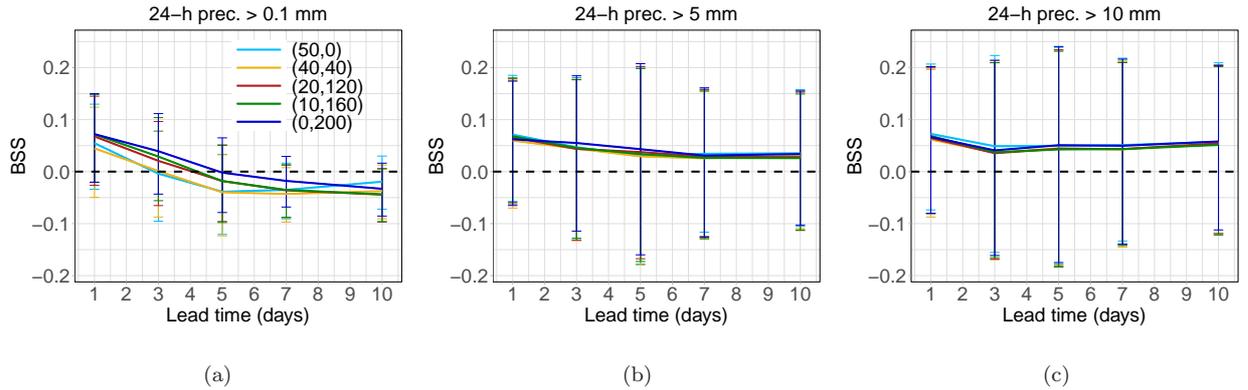}
}
    \centerline{\hbox to 12cm{\scriptsize \qquad (a) \hfill  (b) \hfill (c)}} 
     \caption{Brier Skill scores with respect to the corresponding QM+W forecasts for each dual-resolution combination with 95\,\% confidence intervals for all 3 thresholds.  }
   \label{fig:bssqmw}
\end{figure}

The analysis of Brier skill scores with thresholds 0.1 mm, 5 mm and 10 mm leads us to rather similar conclusions. CSG EMOS forecasts outperform the ECMWF high-resolution (50,0) precipitation accumulation forecast for all lead investigated lead times for all dual-resolution combinations and all three thresholds (Figure \ref{fig:bssraw}); however, for 0.1 mm at day 10 the difference is significant on a 5\,\% level only for the EMOS models based either on the pure high-, or on the pure low-resolution ensemble. Further, Figure \ref{fig:bssqmw} again confirms that there is no dual-resolution combination and forecast horizon where the difference in skill between the matching CSG EMOS and QM+W forecasts  is significant on a 5\,\% level.

The simultaneous DM tests for all considered stations also confirm, that there are no real differences in skill between the CSG EMOS models corresponding to various dual-resolution mixtures. At days 1, 3, 5 and 10 practically there are no stations where the difference in mean CRPS between any pairs of combinations is significant at a 5\,\% level, whereas at day 7 just mixtures (50,0) and (40,40) differ significantly at 1.47\,\%  of the locations. Up to day 7 this is also the only mixture, where for all three thresholds there are stations with significantly different mean BS; however, their proportions vary just between 4.41\,\%  and 6.85\,\% for 0.1 mm, 3.4\,\% and  10.72\,\% for 5 mm and 2.13\,\% and 11.68\,\% for 10 mm. At day 10 the situation changes as for all pairs of mixtures and all thresholds there are locations with significantly different mean BS; though their proportions just barely exceed 9\,\%.

\begin{figure}[t!]
\centering
\hbox{
\includegraphics[width=.33\textwidth]{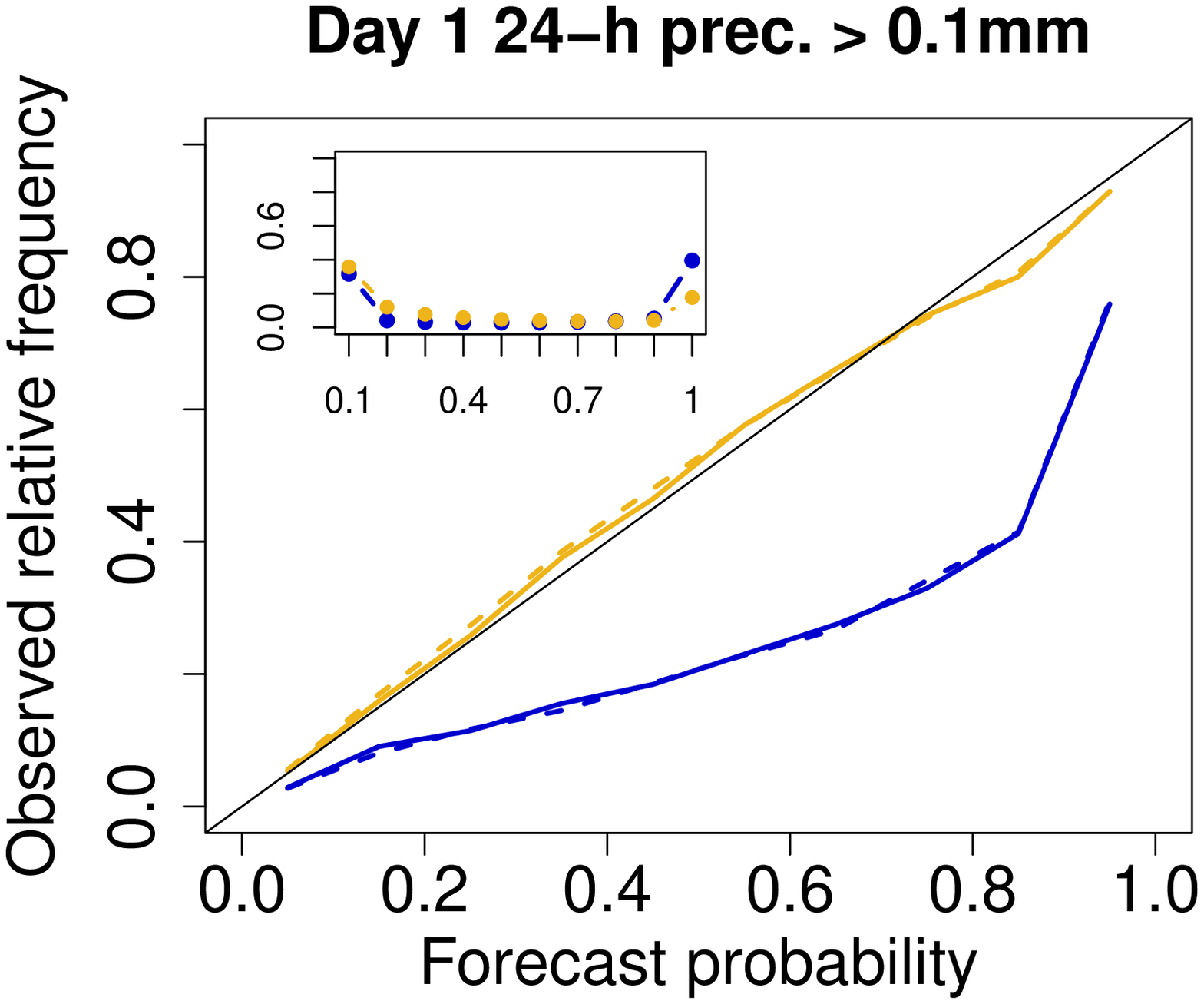} 
\includegraphics[width=.33\textwidth]{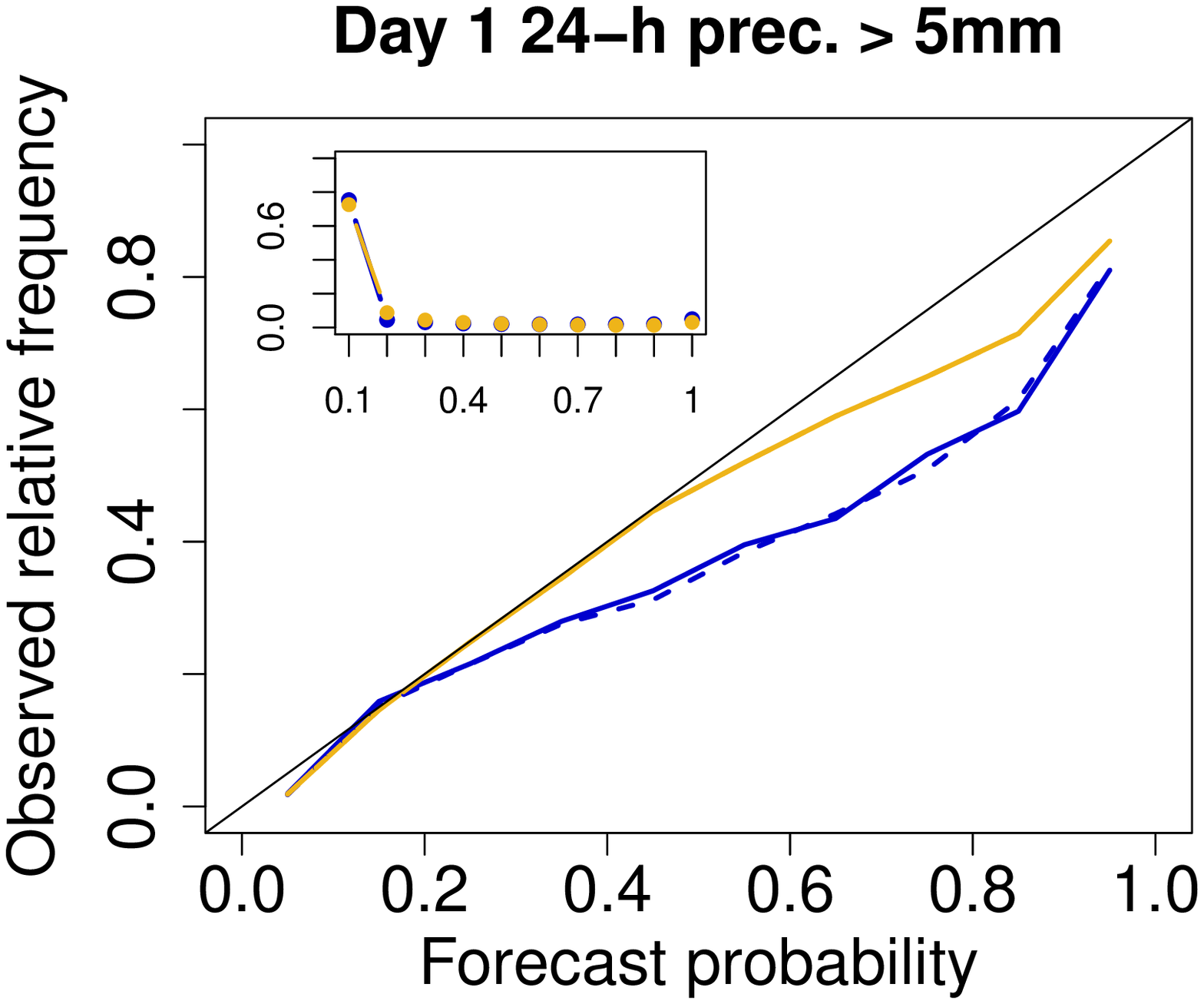}
\includegraphics[width=.33\textwidth]{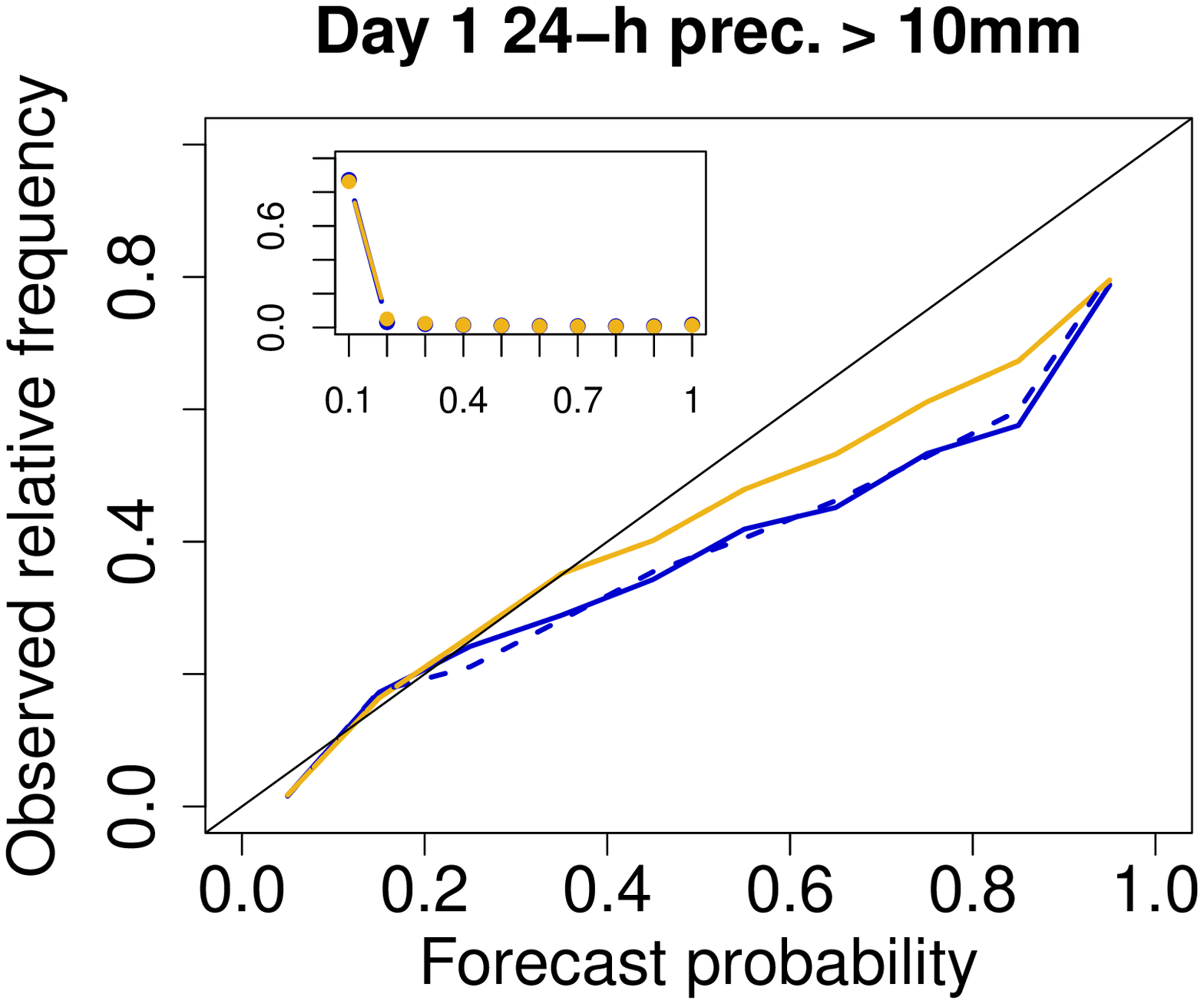}
}
\smallskip
\hbox{
\includegraphics[width=.33\textwidth]{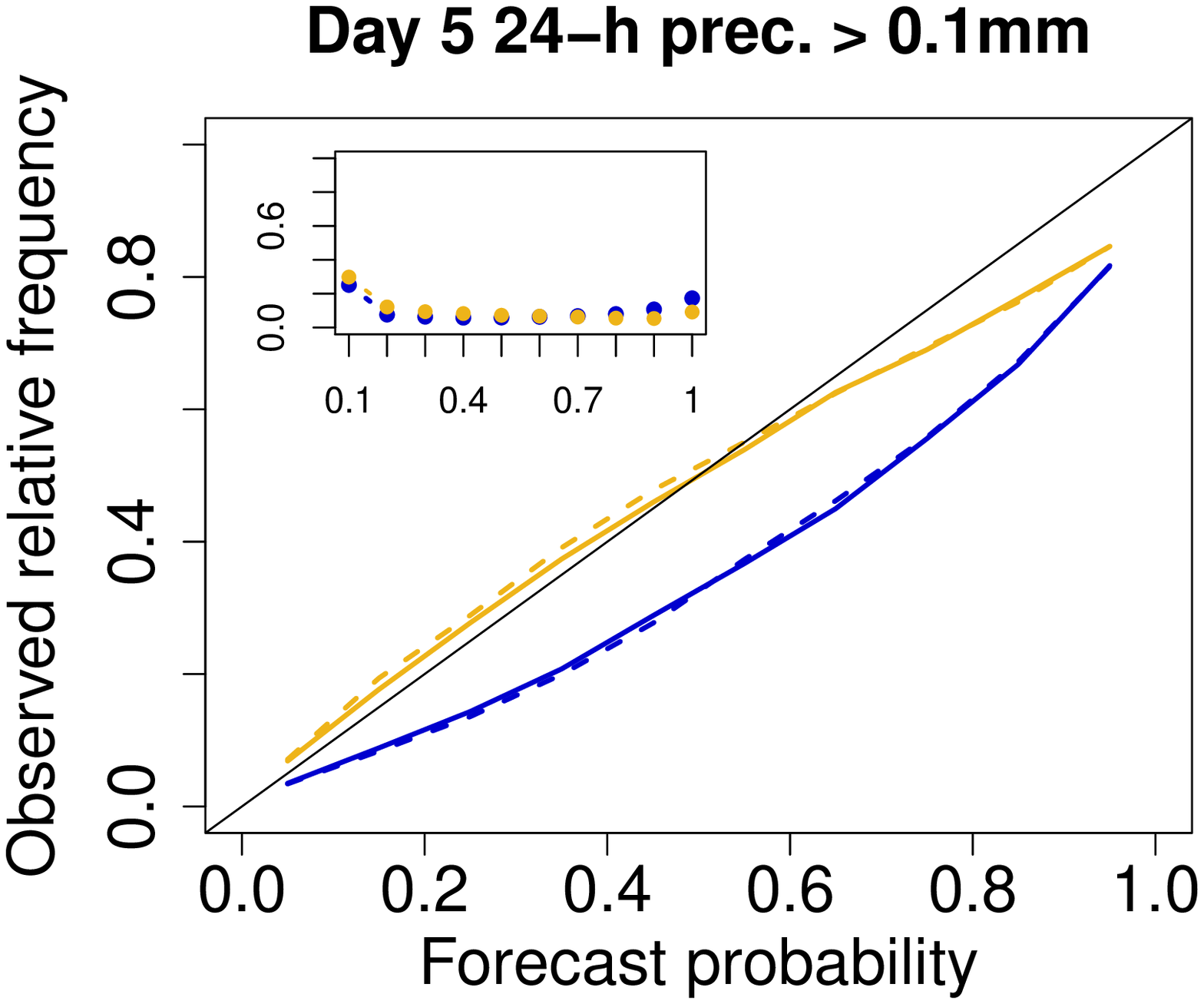} 
\includegraphics[width=.33\textwidth]{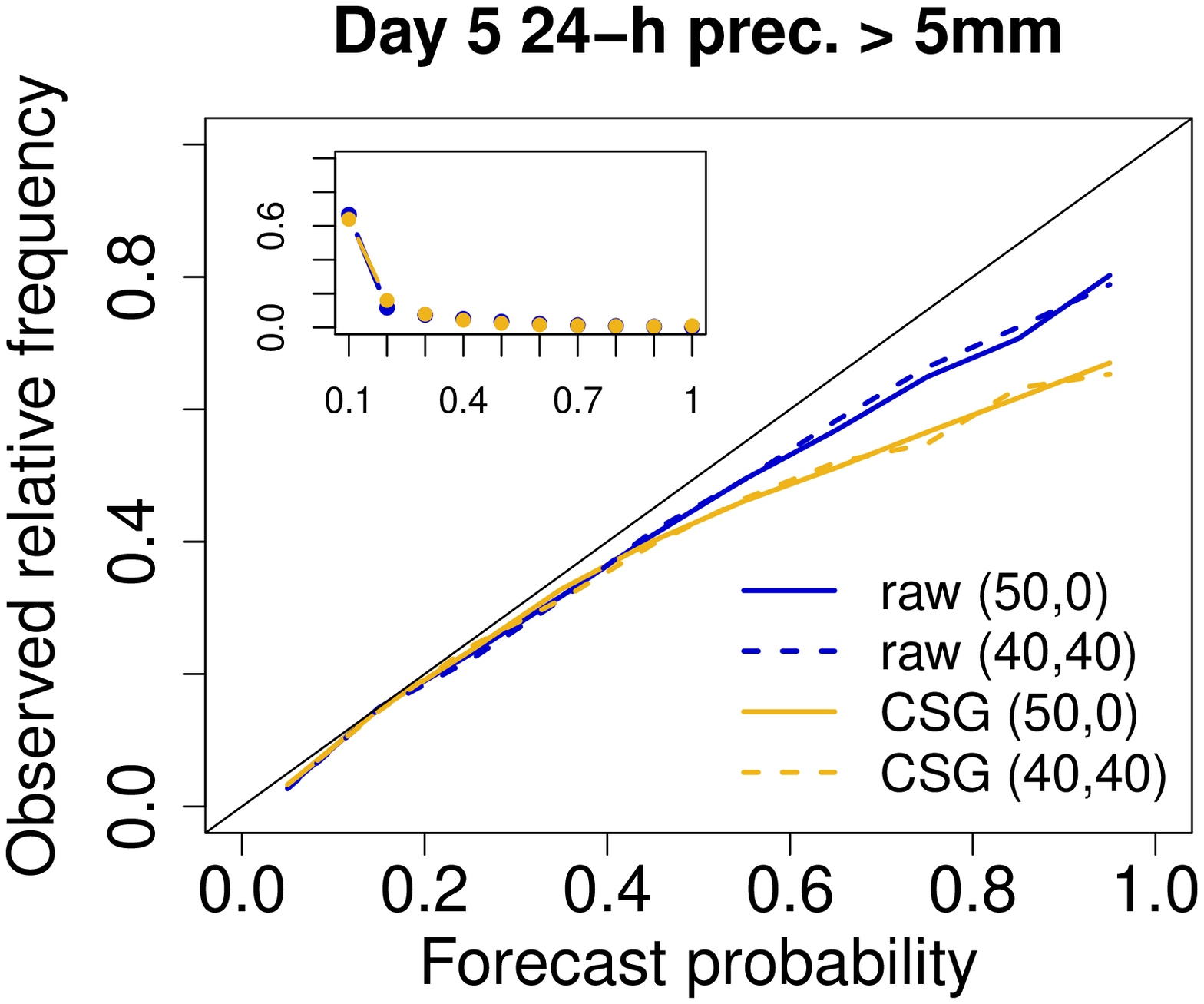}
\includegraphics[width=.33\textwidth]{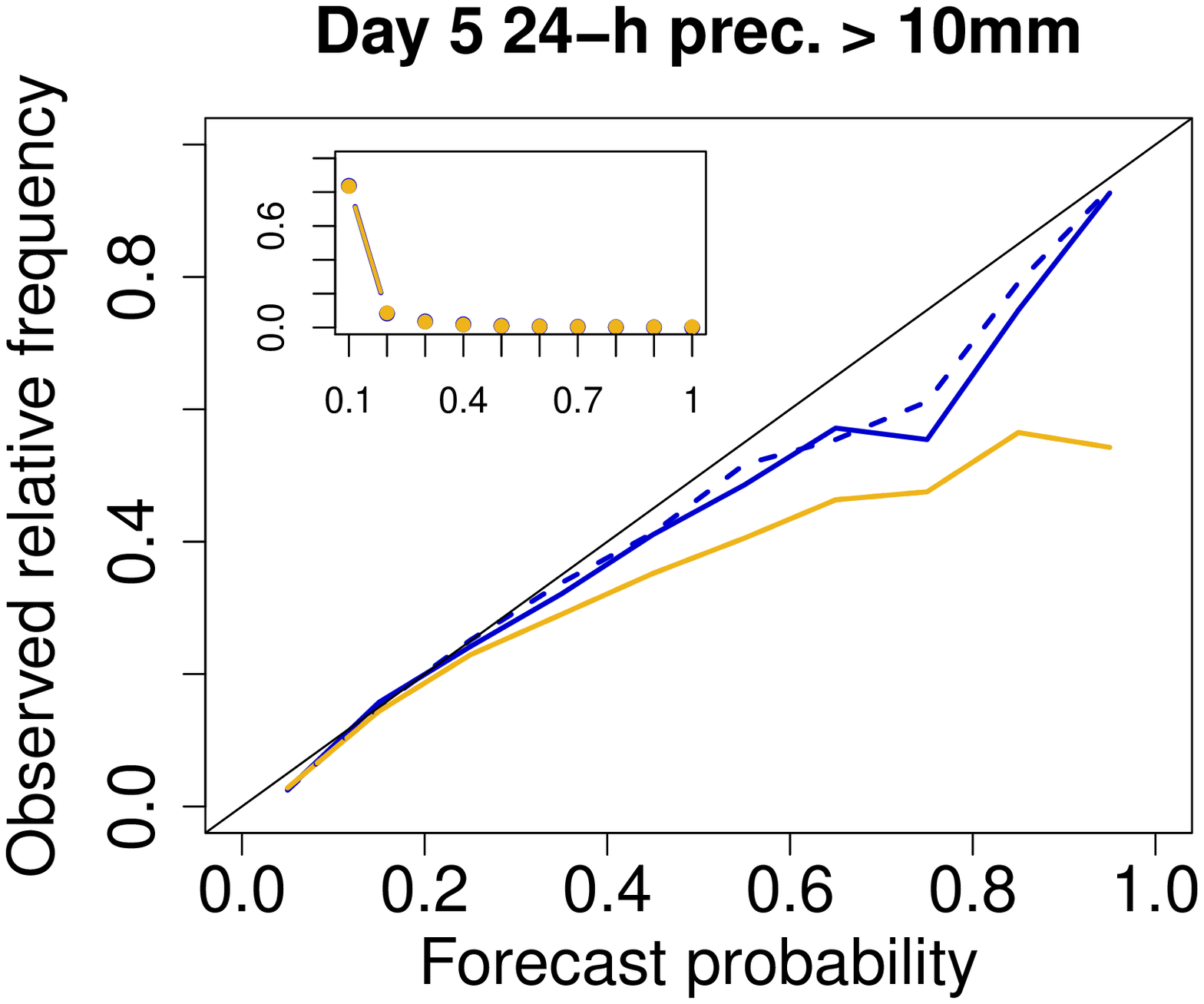}
}
\smallskip
\hbox{
\includegraphics[width=.33\textwidth]{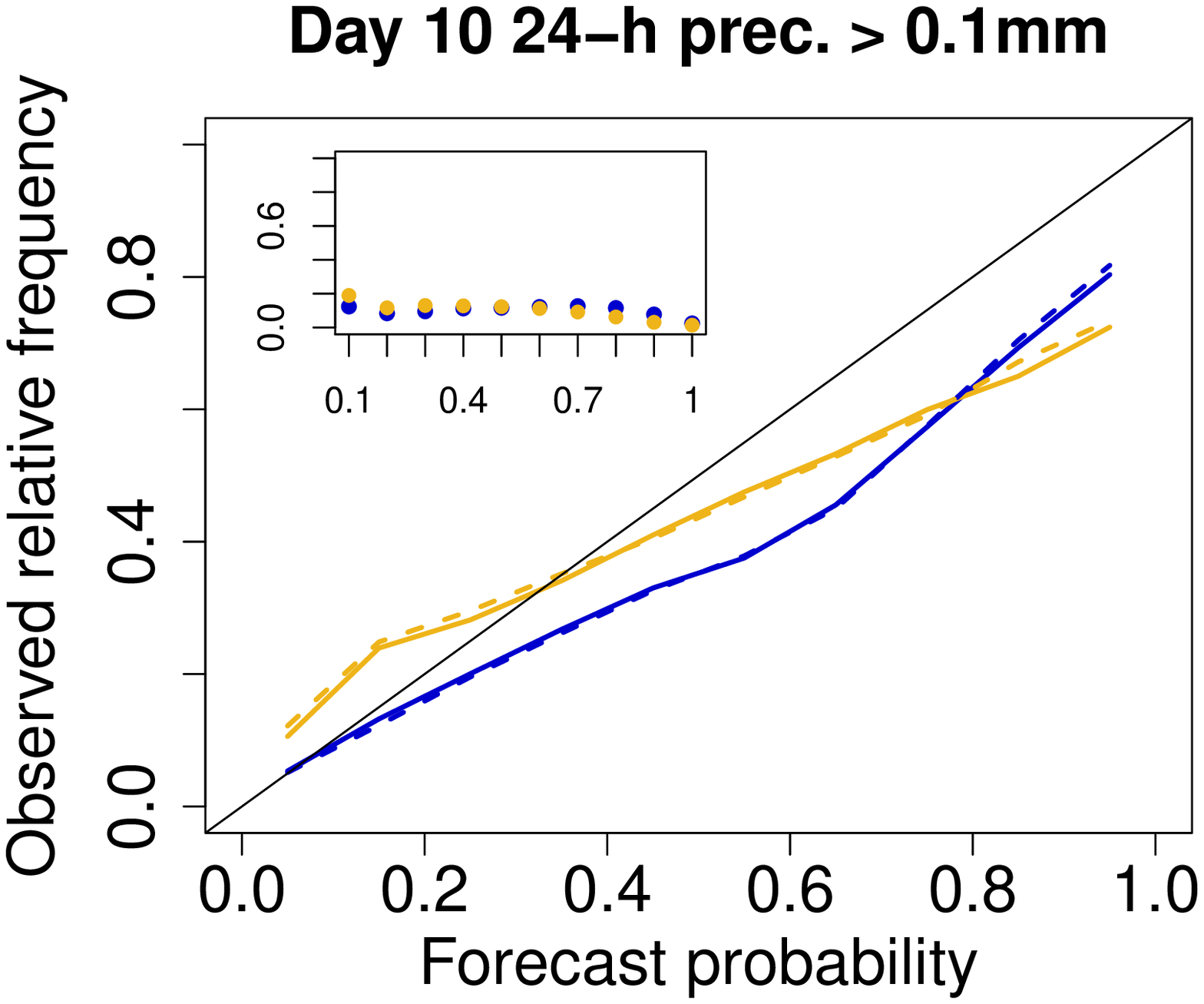} 
\includegraphics[width=.33\textwidth]{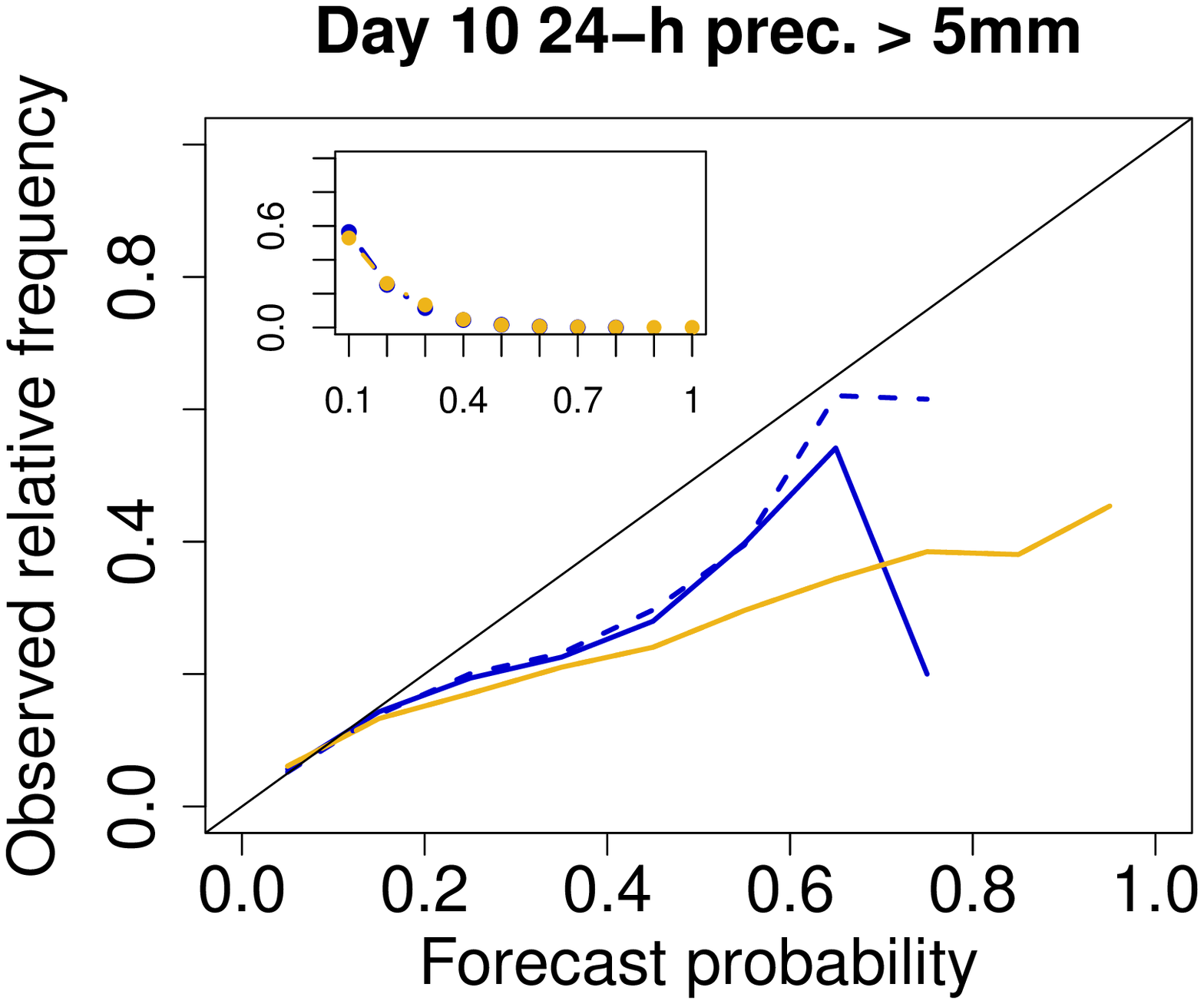}
\includegraphics[width=.33\textwidth]{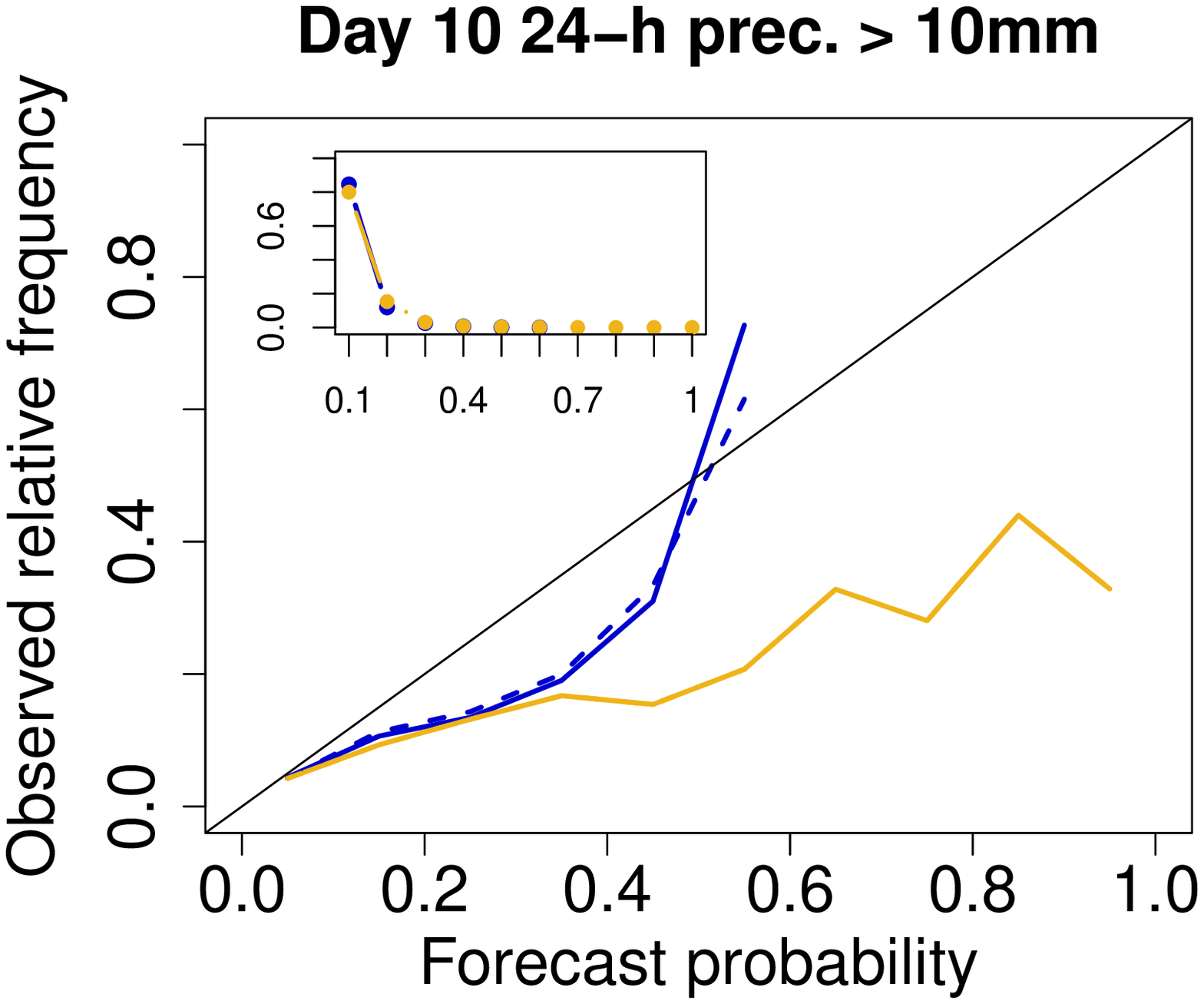}
}
     \caption{Reliability diagrams for 0.1, 5 and 10 mm thresholds of raw (50,0) and (40,40) combinations and corresponding CSG EMOS forecasts for days 1, 5 and 10. The inset curves display the relative frequency of cases within the respective bins for the (50,0) mixture.}
   \label{fig:reldiag}
\end{figure}

Further, Figure \ref{fig:reldiag} displays the reliability diagrams for 0.1, 5 and 10 mm thresholds of raw (50,0) and (40,40) combinations and corresponding CSG EMOS forecasts for days 1, 5 and 10. At day 1 CSG EMOS models definitely outperform the raw forecasts, especially for the 0.1 mm threshold, where the fit to the reference line is almost perfect. For longer lead times the clear advantage of post-processing is preserved only for the lowest threshold where the diagrams are based on 36.5\,\% (day 5) and 34.9\,\% (day 10) of the observations, whereas for 5 mm and 10 mm these proportions are 11.6\,\% and 10.8\,\% and 5.5\,\% and 6.2\,\%.  Further, as the corresponding inset histograms indicate, the distribution of forecast cases is rather biased with very low frequencies at the upper bins. This shortage of data might explain the hectic behaviour of the reliability diagrams for 5 mm and 10mm thresholds at day 10.

\section{Conclusions}
    \label{sec5}

The predictive performance of the censored shifted gamma EMOS approach of \citet{bn16} for statistical post-processing is investigated with the help of various dual-resolution 24h precipitation accumulation ensemble forecasts over Europe. All considered dual-resolution combinations have equal computational cost being equivalent to the cost of the operational 50-member ECMWF ensemble. As reference post-processing approaches we consider the quantile mapping and weighted quantile mapping of \citet{hs18}. All calibration methods are trained using forecast-analysis pairs at EFAS grid points and validated on data of grid points corresponding to SYNOP stations. 

Compared with the raw ensemble combinations, semi-local EMOS post-processing results in a significant improvement for all studied lead times both in terms of the mean CRPS and the mean BS for various thresholds. Moreover, in contrast to the raw ensemble where up to day 5 the mixture of 40 high- and 40 low-resolution forecasts significantly outperforms the other combinations \citep{esti19}, there are no significant differences between the skill of CSG EMOS forecasts corresponding to the various mixtures. Further, in terms of the mean CRPS CSG EMOS forecasts outperform the reference QM and QM+W predictions for all lead times; however, none of the differences are significant. The same is true for the differences between CSG EMOS and QM+W in terms of the Brier scores. These results indicate that the semi-local CSG EMOS method trained merely using data from a 30-day rolling training period is fully able to catch up with the essentially more complex quantile mapping based on historical data of a 20-year period.

The introduction of the new cycle at the ECMWF from 2023, where the current operational setup of a single TCo1279 and 51 TCo639 forecasts will be replaced by 51 forecasts at TCo1279 resolution and 101 forecasts at TCo319 resolution 
immediately provides new avenues of further research on calibration of dual-resolution predictions.  Another possible direction is the investigation of the skill of machine learning-based parametric post-processing approaches in the dual-resolution context, focusing on methods that, similar to EMOS, require short training data, see e.g. \citet{bb21, bb22, gzshf22}.

\bigskip
\noindent
{\bf Acknowledgements.} \ S\'andor Baran and Marianna Szab\'o were supported by the Hungarian National Research, Development and Innovation Office under Grant No. K142849. They are also grateful to the ECMWF for supporting their research stay in Reading. Finally, the authors are indebted to Martin Leutbecher for his help in connection with the new ECMWF IFS configuration.

\end{document}